\documentclass[10pt,preprint2]{aastex}
\shorttitle{Long-Term Variability in NGC 6791}
\shortauthors{Mochejska, Stanek \& Kaluzny}
\begin{document}

\title{Long-term variability survey of the old open cluster NGC 6791.}

\author{B.~J.~Mochejska\altaffilmark{1}, K. Z. Stanek}
\altaffiltext{1}{Hubble Fellow}
\affil{Harvard-Smithsonian Center for Astrophysics, 60 Garden St.,
Cambridge, MA~02138}
\email{bmochejs@cfa.harvard.edu, kstanek@cfa.harvard.edu}
\author{J.~Kaluzny}
\affil{Copernicus Astronomical Center, Bartycka 18, 00-716 Warszawa}
\email{jka@camk.edu.pl}

\begin{abstract}
We present the results of a long-term variability survey of the old
open cluster NGC 6791. The $BVI$ observations, collected over a time
span of 6 years, were analyzed using the ISIS image subtraction
package. The main target of our observations were two cataclysmic
variables B7 and B8. We have identified possible cycle lenghts of
about 25 and 18 days for B7 and B8, respectively. We tentatively
classify B7 as a VY Scl type nova-like variable or a Z Cam type dwarf
nova. B8 is most likely an SS Cygni type dwarf nova. We have also
extracted the light curves of 42 other previously reported variable
stars and discovered seven new ones. The new variables show
long-period or non-periodic variability. The long baseline of our
observations has also allowed us to derive more precise periods for
the variables, especially for the short period eclipsing binaries.
\end{abstract}

\keywords{ open clusters and associations: individual: NGC 6791 --
cataclysmic variables -- binaries: eclipsing -- stars: variables:
other -- color-magnitude diagrams }

\section{{\sc Introduction}}

The open cluster NGC 6791 is unique in many ways. At an age of about 8
Gyr it is believed to be the oldest open cluster in the Galaxy
(Chaboyer et al.\ 1999, Kaluzny \& Rucinski 1995). It is also probably
the most metal-rich, with [Fe/H] estimates ranging from 0.1 to 0.5 dex
(Friel et al.\ 2002, Chaboyer et al.\ 1999, Peterson \& Green 1998,
Kaluzny \& Rucinski 1995). NGC 6791 possesses two of the three
cataclysmic variables (CVs) found in open clusters (Kaluzny et al.\
1997), with the third residing in M67 (Gilliland et al.\
1991)\footnote{ Another unconfirmed as of yet candidate was reported
by Mochejska \& Kaluzny (1999) in NGC 7789.}.

Variability in NGC 6791 was first studied by Kaluzny \& Rucinski
(1993), who discovered 17 variable stars, among them one CV
candidate. Rucinski, Kaluzny \& Hilditch (1996) reported additional 5
variables, including another potential CV. The CV candidates were
confirmed spectroscopically by Kaluzny et al.\ (1997). Mochejska et 
al.\ (2002; hereafter M02) discovered additional 47 variables, 
bringing the total to 69.

In this paper we present a long-term variability study of the open
cluster NGC 6791. Over a time span of nearly 6 years we have collected
465 observations in $V$, 229 in $I$ and 72 in $B$. Our main motivation
was to study the long-term behavior of the two cataclysmic variables.
This dataset also offered us the unique possibility of investigating
long period and non-periodic variables in the cluster.

The paper is organized as follows: Section 2 describes the
observations, \S 3 summarizes the data reduction and variable
selection procedures and \S 4 contains the variable star
catalog. Concluding remarks are found in \S 5.

\section{{\sc Observations}} 
The data analyzed in this paper were obtained between September 1996
and May 2002 at three telescopes, using four CCD cameras.  The
observations were collected during:
\begin{enumerate}
\item
28 nights between 8 Sep and 23 Oct 1996 on the 1.3~m McGraw-Hill
Telescope at the Michigan-Dartmouth-MIT (MDM) Observatory, equipped
with the front-illuminated, Loral $2048^2$ CCD ``Wilbur'' (Metzger,
Tonry \& Luppino 1993). This dataset will hereafter be referred to as 
the $MDM$ dataset.
\item
33 nights between 26 Oct 1996 and 9 Oct 1997 on the 1.2~m telescope at
the F. L. Whipple Observatory (FLWO), equipped with the thinned,
back-illuminated, AR coated Loral $2048^2$ CCD ``AndyCam''
(Szentgyorgyi et al.\ 2002), hereafter $AndyCam$ dataset.
\item
47 nights between 19 Sep 1998 and 22 May 2002 on the 1.2~m telescope
at FLWO, equipped with the ``4Shooter'' CCD mosaic with four thinned,
back-illuminated, AR coated Loral $2048^2$ CCDs (Szentgyorgyi et
al.\ 2002), hereafter $4Shooter$ dataset. Only data from chip 3, centered
on the cluster, were analyzed here.
\item
15 nights between Sep 30 and Nov 7, 1999 and between Oct 14 and 21,
2001 on the the Kitt Peak National Observatory\footnote{Kitt Peak
National Observatory is a division of NOAO, which are operated by the
Association of Universities for Research in Astronomy, Inc. under
cooperative agreement with the National Science Foundation.} (KPNO)
2.1~m telescope, equipped with the Tektronix $2048^2$ CCD ``T2KA'',
hereafter $KPNO$ dataset.
\end{enumerate}

Typical exposure times for the 1~m/2~m telescopes, respectively, were
450/600 s in $B$, 450/300 s in $V$ and 300/300 s in $I$. Some short
exposures were also collected to investigate the variability of bright
stars, which are normally saturated on the longer exposures. Typical
1~m/2~m telescope short exposure times were 45/30 s in $V$ and 30/30 s
in $I$.  Short $B$ exposures were only taken at KPNO and their
exposure times varied between 25 and 180 s.

Over 6 years of monitoring of this cluster we have collected 391 long
and 74 short exposures in $V$, 140/89 in $I$ and 65/7 in $B$. The
number of $BVI$ images in each dataset is listed in
Table~\ref{tab:log}.

\begin{figure}[t]
\plotone{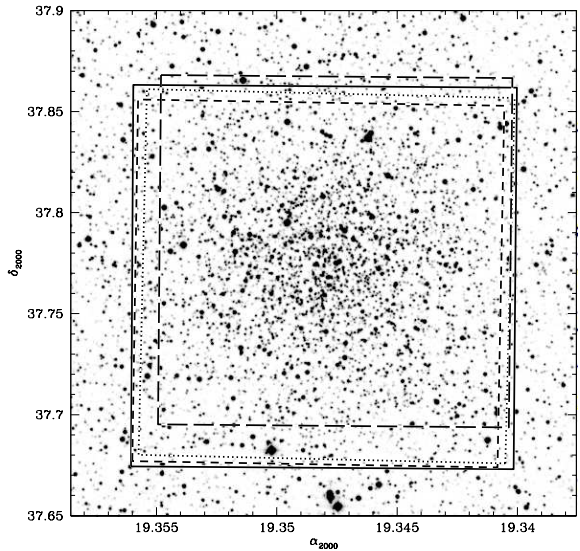}
\caption{Digital Sky Survey image of NGC 6791 showing the field of
view of the 4Shooter chip3 (solid line), MDM (dotted), AndyCam
(short-dashed) and KPNO (long-dashed) CCDs.  North is up and east is
to the left}    
\label{chips}
\end{figure} 

\section{{\sc Data Reduction}}
\subsection{{\it Photometry}}
The preliminary processing of the CCD frames was performed with the
standard routines in the IRAF ccdproc package.\footnote{IRAF is
distributed by the National Optical Astronomy Observatories, which are
operated by the Association of Universities for Research in Astronomy,
Inc., under cooperative agreement with the NSF.}
The KPNO data were corrected for CCD non-linearity at this stage, as
described by Mochejska et al.\ (2001).

Photometry was extracted using the ISIS image subtraction package
(Alard \& Lupton 1998, Alard 2000). A brief outline of the applied
reduction procedure is presented here. For a more detailed description
the reader is referred to M02.

The ISIS reduction procedure consists of the following steps: (1)
transformation of all frames to a common $(x,y)$ coordinate grid; (2)
construction of a reference image from several best exposures; (3)
subtraction of each frame from the reference image; (4) selection of
stars to be photometered and (5) extraction of profile photometry from
the subtracted images.

In further discussion we will refer to images called ``template'' and
``reference''. By ``template'' we mean a single best quality exposure
in a filter and dataset combination. A reference frame is a high S/N
image constructed from the template and several other high quality
exposures processed to match the template point-spread function (PSF)
and background level. A template is chosen and a reference image
constructed for each of the 19 filter and dataset combinations.

All computations were performed with the frames internally subdivided
into four sections ({\tt sub\_x=sub\_y=2}). Differential brightness
variations of the background were fit with a first degree polynomial
({\tt deg\_bg=1}). A convolution kernel varying quadratically with
position was used ({\tt deg\_spatial=2}). The psf width ({\tt
psf\_width}) was set to 33 pixels for all chips, except for the KPNO
data, where it was set to 15 pixels. We used a photometric
radius ({\tt radphot}) of 5 pixels for 4Shooter and AndyCam, 4 pixels
for MDM and 3 pixels for KPNO data.

\subsection{{\it Calibration}}
\begin{figure}[t]
\plotone{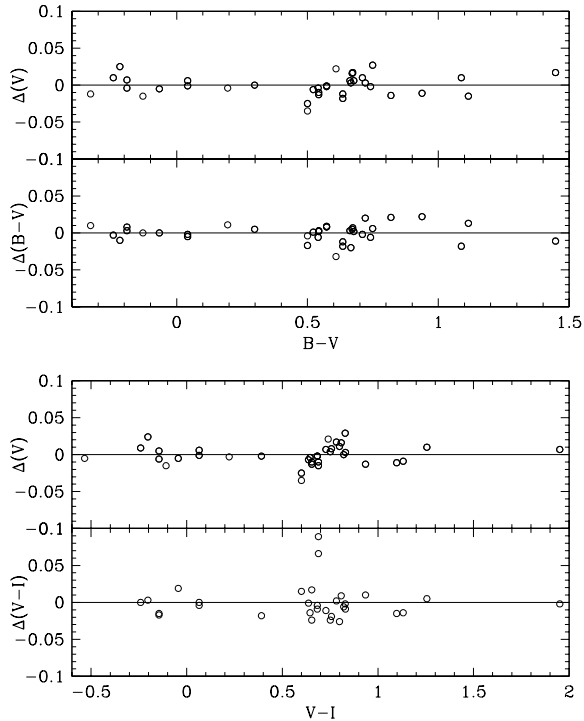}
\caption{$V$, $B-V$ and $V-I$ residuals of Landolt (1992) standard
stars as a function of color for the adopted transformation
equations.}
\label{fig:res}
\end{figure}

The transformations of instrumental magnitudes to the standard system
were derived from observations of 67 stars in 21 Landolt (1992)
standard fields, collected with the 4Shooter on 4 November 1999.
The following transformations were adopted:
\begin{eqnarray*}
\label{eq:vbv}
v = V + 2.1326 + 0.0439 (B-V)+ 0.1480 (X-1.25)\\
\label{eq:bv}
b-v =   0.1275 + 0.9151 (B-V)+ 0.1074 (X-1.25)\\  
\label{eq:vvi}
v = V + 2.1318 + 0.0385 (V-I)+ 0.1478 (X-1.25)\\
\label{eq:vi}
v-i =  -0.7390 + 1.0240 (V-I)+ 0.0880 (X-1.25)
\end{eqnarray*}
Figure \ref{fig:res} shows the $V$, $B-V$ and $V-I$ transformation
residuals of the standard stars as a function of color. The calibrated
4Shooter reference image photometry is shown in V/V-I and V/B-V
color-magnitude diagrams (CMDs) in Figure~\ref{fig:cmd}.

The zero points of the light curves of variable stars from the other
datasets were adjusted to the 4Shooter data by adding an offset. It
was computed as the mean offset for stars within a radius of 300
pixels around the variable, with the rejection of $>3$ sigma
outliers. The color term was not taken into account.

Additional offsets, if necessary, were applied between overlapping
datasets for all variables, ie. MDM-FLWO and 4Shooter-KPNO, and
between all datasets in case of periodic variables. The offsets are
probably due in large part to the difference in color terms between
the chips. Such offsets were added in 14\% of the cases and typically
were $\leq 0.06$ mag. For some variables the offsets were large (up to
0.5 mag) because of blending with nearby stars in the datasets with
inferior seeing.

\subsection{{\it Astrometry}}
The transformation from rectangular to equatorial coordinates was
derived using 997 transformation stars from the USNO A-2 catalog
(Monet et al.\ 1998) for the 4Shooter $V$-band template. The average
difference between the catalog and the computed coordinates for the
transformation stars was $0\farcs 14$ in right ascension and $0\farcs
12$ in declination.

\subsection{{\it Variability Search}}

We used the standard ISIS procedure to search for variable stars. For
every pixel it computes a median of absolute deviations on all
subtracted images and performs a simple rejection of cosmic rays and
defects. The result is stored as an image, where the variables are
then identified as bright peaks. We used this method to search for new
variables in all filter and dataset combinations.

We also extracted the light curves for all stars detected by DAOphot
(Stetson 1987) on the template frames and searched them for
variability using the index $J$ (Stetson 1996), as described in more
detail by M02.

To search for periodicity we used the analysis of variance statistic
(Schwarzenberg-Czerny 1996).

\begin{figure*}[t]
\epsscale{2.17}
\plotone{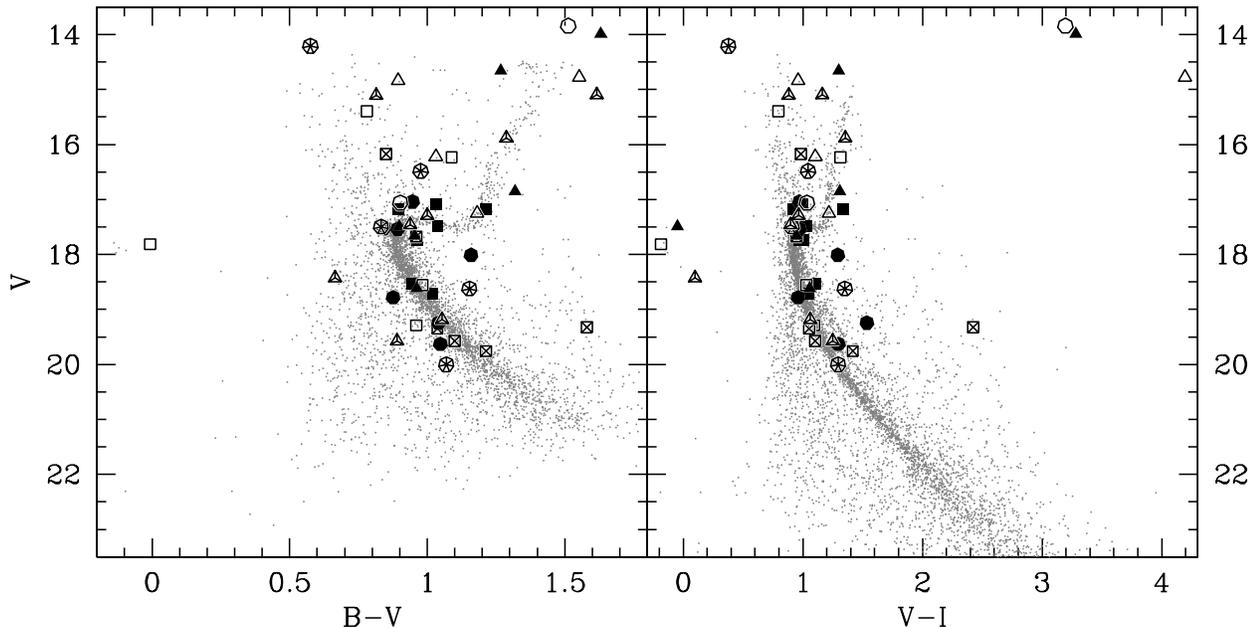}
\epsscale{1}
\caption{The V/V-I and V/B-V color-magnitude diagrams. Non-periodic
variables are indicated by triangles, eclipsing binaries by squares
and periodic variables by circles. Filled symbols indicate membership
probability above 75\%, open symbols -- below 75\%, and skeletal -- no
membership probability estimate.}
\label{fig:cmd}
\end{figure*}

\begin{figure}[t]
\epsscale{1}
\plotone{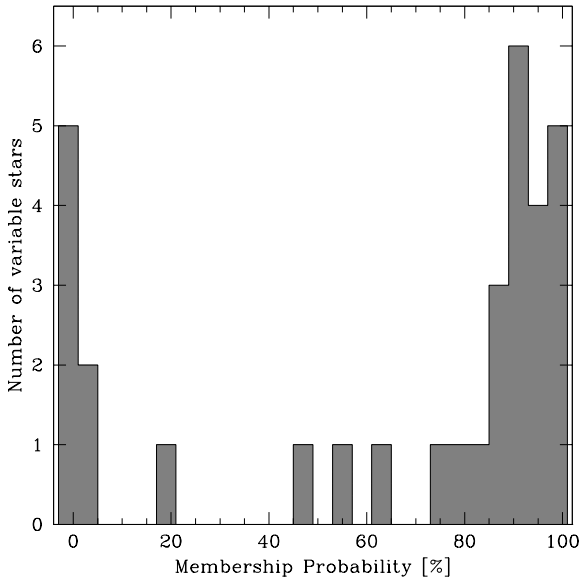}
\caption{The distribution of preliminary cluster membership
probabilities for 32 out of 51 variable stars, kindly supplied by  
Dr.~Kyle Cudworth.}
\label{pro}
\end{figure}

\begin{figure*}[tp]
\epsscale{2}
\plotone{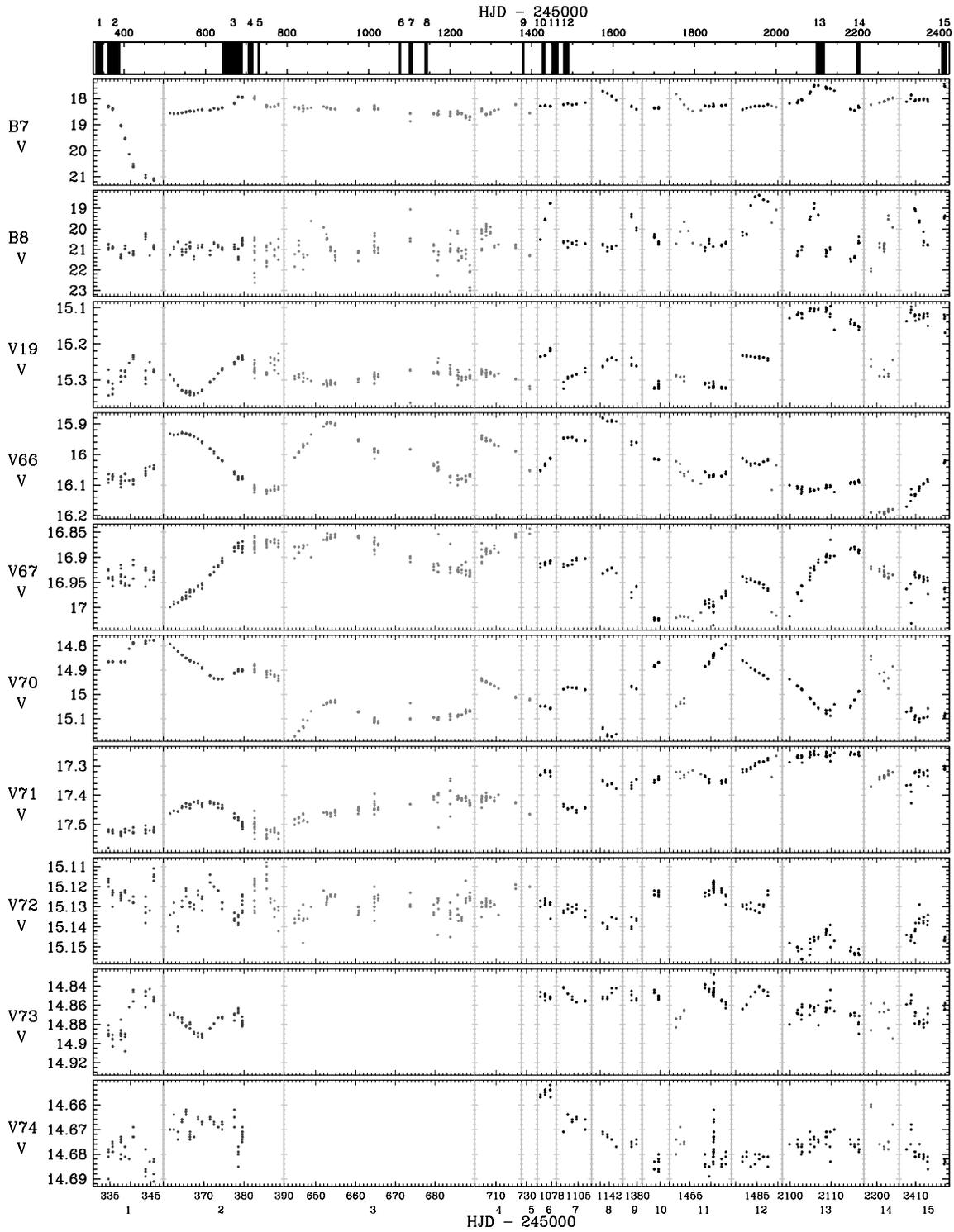}
\epsscale{1}
\label{lcm}
\caption{$V$-band light curves for 10 selected long period and
non-periodic variable stars.  Each dataset is plotted with a different
shade of grey. The top window illustrates the distribution in time of
the 15 sub-windows plotted for the variables.}
\end{figure*}

\begin{figure*}[t] 
\epsscale{2.17}
\plotone{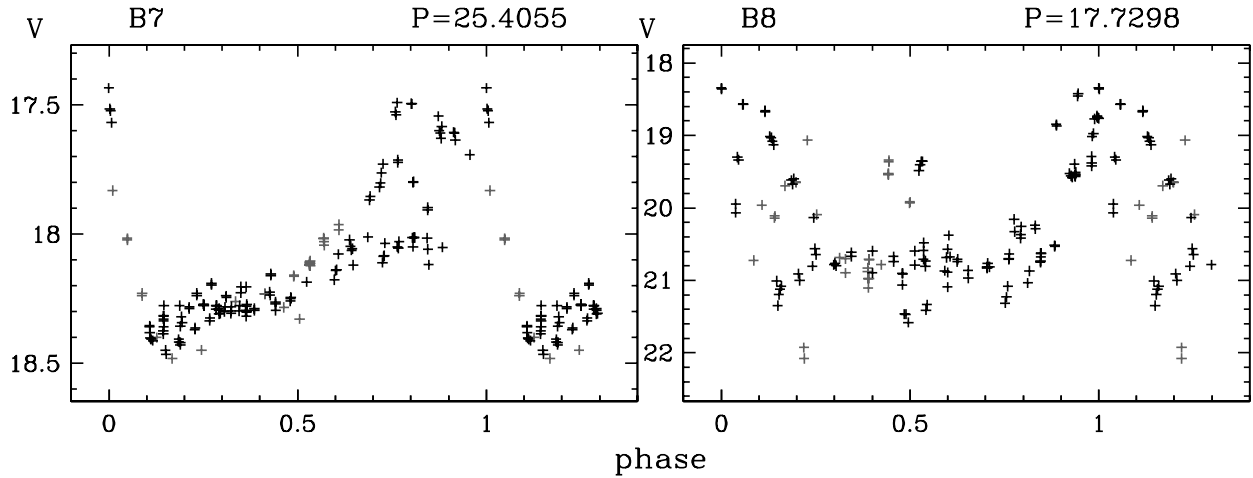}
\epsscale{1}
\caption{Phased $V$-band light curves for the cataclysmic variables
B7 and B8 (4Shooter and KPNO data only).}
\label{lcv}
\end{figure*}

\begin{figure*}[t]
\epsscale{2.17}
\plotone{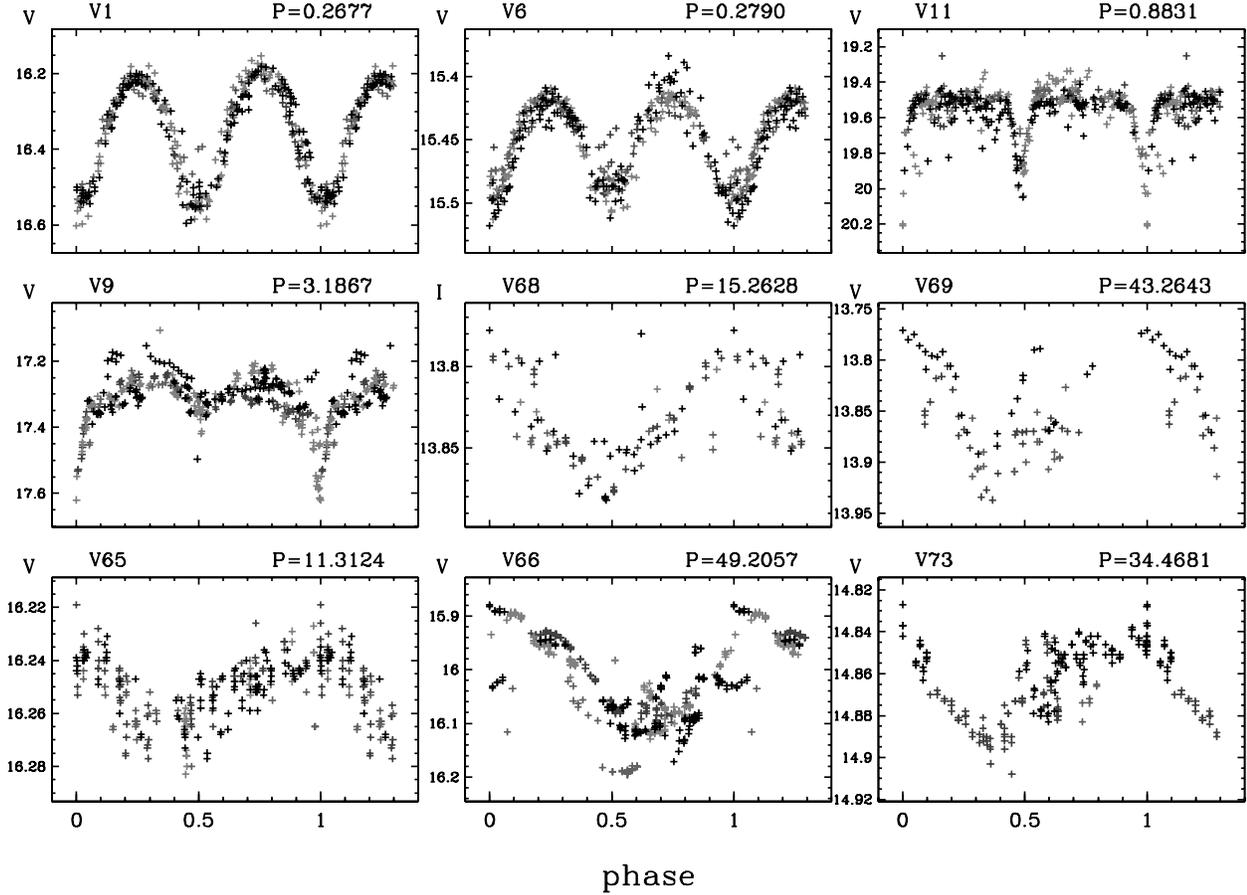}
\epsscale{1}   
\caption{Light curves for six selected eclipsing binaries (top two
rows) and for three variables classified as long-period (bottom
row). Each dataset is plotted with a different shade of grey. The  
light curves are in the $V$-band, with the exception of V68, which is
in the $I$-band.}
\label{lcp}  
\end{figure*}

\begin{figure*}[t]
\epsscale{2.17}
\plotone{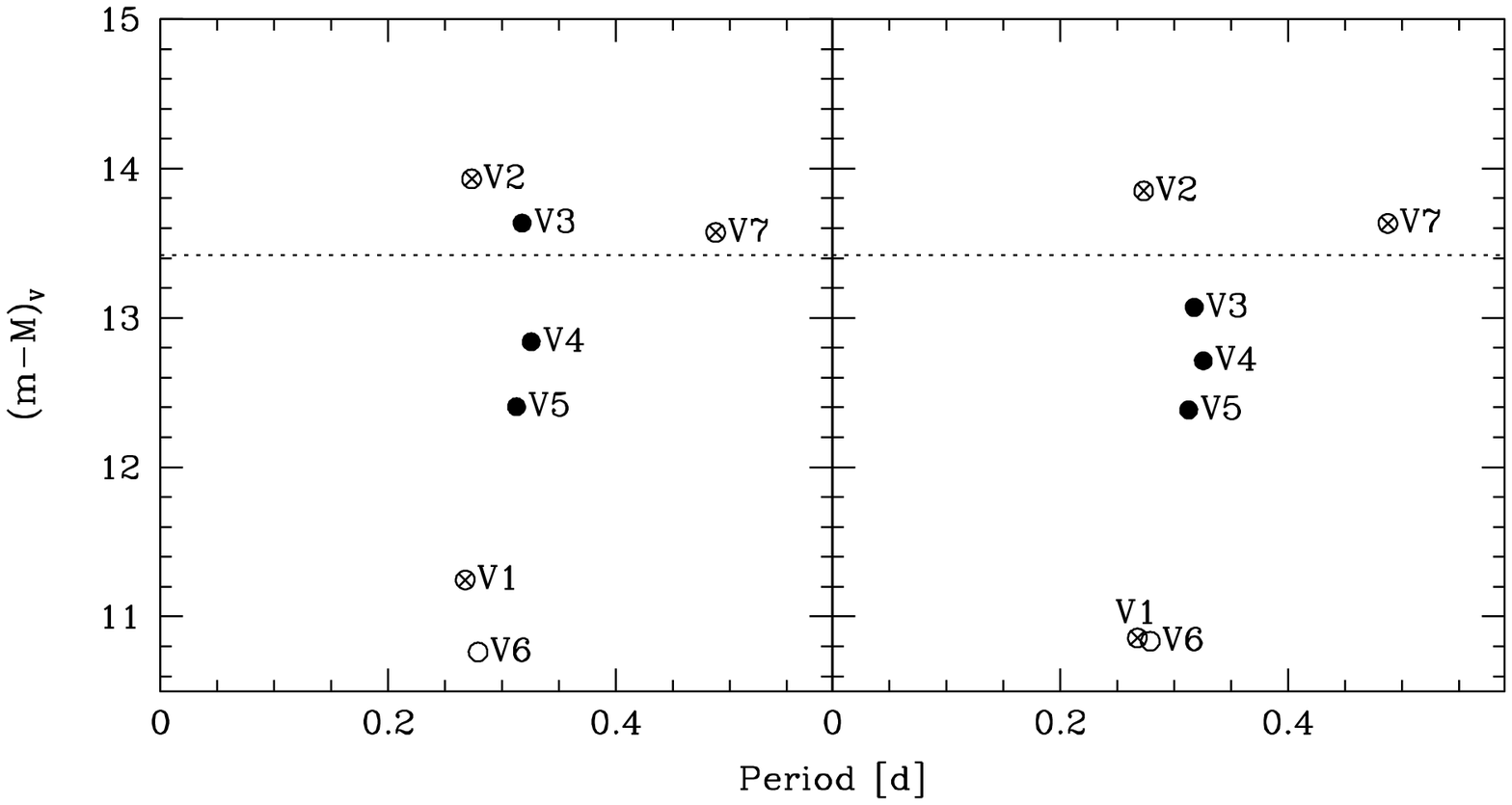}
\epsscale{1}
\caption{Apparent distance modulus as a function of period for the
contact binaries V1-V7. The absolute $V$ magnitude was derived from
the calibration of Rucinski (2000), based on B-V and V-I colors (left
and right panel, respectively). Filled circles indicate membership
probability above 75\%, open symbols -- below 75\%, and skeletal -- no
membership probability estimate.}
\label{WUMa}
\end{figure*}

\section{{\sc Variable Star Catalog}}

We have confirmed the variability of 37 out of 44 previously known
variables within our field of view. Of the 7 variables not recovered,
three are detached eclipsing binaries with no eclipses observed by
us. Many of the variables (V29-V67) were discovered in higher S/N 900
s $R$-band observations (M02), hence it was anticipated that some
fainter and/or lower amplitude objects might not be recovered.  We
have also discovered seven new variables.

The properties of the 51 variable stars are discussed in subsections
4.1 - 4.5.  Tables~\ref{tab:misc}-\ref{tab:pul} list the following
parameters for the variables: the identification number, right
ascension, declination, period and its error estimate (for eclipsing
and periodic variables), $V$, $I$ and $B$-band magnitude (maximum for
non-periodic and eclipsing variables, flux weighted mean for
periodic), variability amplitudes in each band (semiamplitudes for
periodic variables) and preliminary cluster membership probabilities,
kindly provided to us by Dr.~Kyle Cudworth. Table~\ref{tab:misc}
contains an additional column with comments on the variability type.

The variables are plotted on the CMD in Figure~\ref{fig:cmd}.
Non-periodic variables are indicated by triangles, eclipsing binaries
by squares and other periodic variables by circles. Filled symbols
indicate membership probability above 75\%, open symbols -- below
75\%, and skeletal -- no membership probability estimate. The light
curves, mostly in $V$, for a selection of variables are shown in
Figures~\ref{lcp}-\ref{lcm}. The $BVI$ light curves and finder charts
for all variables and machine-readable versions of tables
~\ref{tab:misc}-\ref{tab:pul} are available for download via ftp from
{\tt cfa-ftp.harvard.edu} in the {\tt /pub/bmochejs/NGC6791} directory.

The distribution of the cluster membership probabilities for the
variables is shown in Figure~\ref{pro}. Proper motion data, available
for 32 out of 51 variables, are in most cases conclusive in resolving
the question of their cluster membership. Of those variables, 21 are
likely cluster members and 8 are not, with probabilities in excess of
75\% and below 25\%, respectively. There are only three variables for
which the membership is very uncertain ($25\%<P<75\%$).

\subsection{\it Cataclysmic variables}
Our main interest was to study the long-term behavior of the two
cataclysmic variables B7 (a.k.a.\ V15) and B8. These are unique
objects, as only three such stars have been found and confirmed in all
open clusters. B7 is a cluster member, as shown by proper motion data,
while for B8 no estimate is available. Both were spectroscopically
confirmed to be CVs by Kaluzny et al.\ (1997).

During our monitoring the brighter cataclysmic variable B7 underwent a
large drop in brightness, by 3 mag in $V$ and 2 mag in $I$, and three
outbursts of about 0.5-1 mag. Smaller amplitude variations, as well as
long-term trends are also present in the light curve. 

The large decline in the MDM data and intermediate brightness at other
times would suggest that B7 may be an VY Scl type nova-like (NL)
variable, which vary little about their mean magnitude, but
occasionally fall in brightness by $>1$ mag (Warner 1995). Another
possibility is that it is a Z Cam type dwarf nova (DN), which exhibit
periods of standstill at irregular intervals. One notable difference
between these two types of CVs is that Z Cam DNe are brighter by 0.5-1
mag in outburst than in the standstill phase, while the apparent
outbursts in VY Scl type NL variables reach at most the high state
brightness. We have observed the variable at minimum light only once,
when it fell from the mean brightness by 3 magnitudes, so we are
unable to classify B7 based on this criterion. The 4Shooter and KPNO
data phase rather well with a period of $\sim25.4$ days, which is
compatible with the typical cycle length of 10-30 days for a Z Cam
type DN (Sterken \& Jaschek 1996).

The other cataclysmic variable, B8, displayed several large outbursts
about 3 magnitudes in amplitude, as well as a few smaller ones. The
MDM and AndyCam data display incoherent variability, possibly due to
flickering. 

The variability of B8 is reminiscent of an SS Cygni type dwarf nova
(Sterken \& Jaschek 1996).  The 4Shooter and KPNO light curve phases
reasonably with a period of $\sim17.73$ days, which if real, is
shorter than the typical range of 30-100 days for these variables.

\subsection{\it Newly Discovered Variables}
We have identified a total of seven new variables in the field of
NGC 6791.

V68, located 4 magnitudes above and 0.3 mag to the blue of the
turnoff, is a periodic variable, with a period of $\sim15$ days.
It may be an RV Tau variable with twice the period. 

V69 and V70 are very red, bright variables. V69 seems to be periodic,
with a period of about 43 or 86 days. The period determinations are
very uncertain and different for the $V$ and $B$ bands due to the
star's long period and sparse sampling -- the star was unsaturated
only on some of the short exposures. V70 is an irregular variable with
an amplitude of $\sim0.4$ mag in $V$. The preliminary proper motion
data are inconclusive as to the cluster membership of these variables,
yielding probabilities of 47\% for V69 and 62\% for V70.

V71 is also irregular, with a $V$-band amplitude of $\sim0.3$ mag. On
the CMD it is located near the base of RGB. Proper motion data gives a
membership probability of 55\% for this variable.

V72 and V73 are located about 2 mag above the cluster turnoff. They
exhibit irregular variability with $V$-band amplitudes of about
$0.04$ and $0.08$ mag, respectively. When the observations between 
HJD 2452099 and 2452204 are excluded, the $V$-band light curve for 
V73 can be phased with a period of $\sim34$ days (Figure~\ref{lcp}). 
Its membership probability is low (20\%).

V74 exhibits irregular brightness variations with an amplitude of
$0.04$ mag in $V$. On the CMD it is located just blueward of the red
clump. It is very likely a cluster member.

\subsection{\it Previously known long period and non-periodic variables}

During previous investigations the cluster was found to possess six
other long period and non-periodic variables: V13, V19, V62 and V65-V67.
V13 and V19 show irregular variability with sudden rises in brightness
by 0.1-0.2 mag in $V$. V13 is a cluster member.  We confirm the
variability of V62, but due to its faintness and poor quality light
curve, we are unable to fathom its nature. V65 seems to be  periodic,
with a period of $\sim 11.3$ days, when the MDM dataset and May 2002
4Shooter run are excluded (Figure~\ref{lcp}). It does not belong to the
cluster.  Of the two variables on the RGB, V66 and V67 (confirmed
member), the first is more or less periodic, with a period of $\sim49$
or $\sim99$ days (Figure~\ref{lcp}), while the second seems to be
irregular.

Of the five variables classified in previous investigations as
detached eclipsing binaries with unknown periods, V10, V18, V20, V21
and V60, we have detected eclipses only in V18 and V20, but were
unable to determine convincing periods. V10 seems to have decreased
its brightness by 0.1 mag between the 1999 and 2001 seasons. V60 also
displays a long-term trend, especially visible in the 4Shooter data.
Based on proper motions, V18 and V60 are most likely cluster
members.

\subsection{\it Previously known eclipsing binaries}

Thanks to the long time span of our observations, we have derived more
precise periods for the eclipsing binaries B4, V1, V3-V7, V9, V11,
V12, V16 and V38. We have confirmed the variability of V2 and V14 with
the previously reported periods, but were unable to improve their
determinations. Variables V3-V5, V9, V16, V38, V12 are likely cluster
members, while B4, V6 and V14 are non-members. 

Figure \ref{WUMa} shows a plot of apparent distance modulus $(m-M)_V$ as
a function of period for W UMa type contact binaries V1-V7. The absolute
$V$ magnitude was derived from the calibration of Rucinski (2000), based
on B-V and V-I colors (left and right panel, respectively). A reddening
of E(B-V)=0.1 (Chaboyer et al.\ 1999) was adopted. Filled circles
indicate membership probability above 75\%, open symbols -- below 75\%,
and skeletal -- no membership probability estimate. Of the three
variables, which are high probability cluster members based on proper
motions, V3 and V4 are within 0.7 mag of the cluster distance modulus
$(m-M)_V = 13.42$ (dotted line; Chaboyer et al.\ 1999) and V5 is about a
magnitude brighter than predicted from the calibration. V6, which is a
non-member, based on its proper motion, is located $\sim2.5$ magnitudes
in front of the cluster. Of the three variables with no membership
probability estimate, V2 and V7 are probably located at the distance of
the cluster, while V1 seems closer by $\sim2$ magnitudes.

V29, which was classified in M02 as an eclipsing binary with a
period of $\sim0.4$ days, shows a long-term rise in brightness with an
amplitude of $\sim0.5$ mag during our observations (the large
amplitude in Table~\ref{tab:ecl} is mainly due to scatter in the
AndyCam data).  The eclipses were not recovered.

For V31 and V32 we did not detect any coherent variability at the
periods determined in M02. These periods are also not confirmed by the
additional extensive $R$-band photometry for this cluster, collected
by the authors (Mochejska et al.~2003, in preparation), where the
variables only show weak long-term trends. V31 is a probable cluster
member and V32 is most likely not.

V33, previously classified as an EB with a period of $\sim2.37$ days,
exhibits long term brightness variations. In the 4Shooter data we have
observed a recovery by 0.15 mag to its maximum observed brightness. It
also displays quasi-periodic variations with roughly the period
reported in M02. This variable is most likely not a member of the
cluster.

The light curve for V37 is very noisy and we were not able to recover
the previously determined period. In the M02 data the variable
underwent a 0.1 mag rise in brightness. It appears to have experienced
a rapid increase in brightness by 0.6 mag at the beginning of the
AndyCam run.

\subsection{\it Previously Known Other Periodic Variables}

We have confirmed the previously determined periods of V41, V48, V54,
V56 and V58. We have confirmed the variability of V17, V45, V46 and
V52 with previously derived periods, but were unable to improve their
determinations. V45 seems to exhibit long term brightness variations,
besides short term variability. We were unable to confirm the periods
of V42, V51 and V53, determined by M02. V42 seems to have steadily
decreased its brightness by $\sim0.6$ in $V$. Variables V17, V41, V42,
V48, V53, V56, V58 are high probability cluster members, and V45 is
most likely a non-member.

Based on its period and location on the color-magnitude diagrams, V17
might be a member of the newly proposed class of variable stars termed
``red stragglers'' (Albrow et al.\ 2001) or ``sub-subgiant stars''
(Mathieu et al.\ 2003). To date, six such stars have been found in 47
Tuc (Albrow et al.\ 2001) and two in M67 (Mathieu et al.\ 2003). Thus
far, the origin and evolutionary status of these stars remains
unknown.

\section{{\sc Conclusions}}

In this paper we have presented $BVI$ photometry for 51 variable stars
in the field of the open cluster NGC 6791. Over 6 years of monitoring
of this cluster we have collected 391 long and 74 short exposures in
$V$, 140/89 in $I$ and 65/7 in $B$. The light curves of the variables
were extracted using the ISIS image subtraction package (Alard \&
Lupton 1998).

The main target of our observations were the two cataclysmic variables
B7 and B8. Their light curves show outbursts characteristic of CVs. We
have identified possible cycle lenghts of 25 and 18 days for B7 and
B8, respectively. We tentatively classify B7 as a VY Scl type
nova-like variable or a Z Cam type dwarf nova. More observations of
the variable at minimum brightness would help elucidate its nature. B8
is most likely an SS Cygni type dwarf nova.

We have also discovered seven new variables: two periodic and five
long-period or non-periodic ones. In addition, we have extracted the
light curves of 42 other previously reported variable stars. For 31 of
them we confirm the variability period, if applicable, and type, given
by M02 and for 17 we have derived new, more precise periods.

\acknowledgments{ We would like to thank G\'asp\'ar Bakos, David
Bersier, Martin Krockenberger, Lucas Macri, Dimitar Sasselov and Andy
Szentgyorgyi for their help in obtaining the data and Alex
Schwarzenberg-Czerny for software to compute errors in the period
determinations. We also thank Dr.~Kyle Cudworth for kindly supplying
us with his preliminary cluster membership data for the variables. We
would also like to thank the referee, Ronald Gilliland, for a prompt
and useful report, which significantly improved the paper. BJM was
supported by the Polish KBN grant 5P03D004.21 and the Foundation for
Polish Science stipend for young scientists. Support for BJM was
provided by NASA through Hubble Fellowship grant HST-HF-01155.01-A
from the Space Telescope Science Institute, which is operated by the
Association of Universities for Research in Astronomy, Incorporated,
under NASA contract NAS5-26555.}

\clearpage
\begin{deluxetable}{llrrrrrrr}
\tabletypesize{\footnotesize}
\tablewidth{0pt}
\tablecaption{Log of observations for NGC 6791}
\tablehead{\colhead{Telescope} & \colhead{CCD}& 
\colhead{$\arcsec$/pix}& \colhead{B} & \colhead{B$_{sh}$}&
\colhead{V}& \colhead{V$_{sh}$}& \colhead{I}& \colhead{I$_{sh}$}}
\startdata
MDM  1.3m & Wilbur   & 0.32 &  - &  - &  79 & 29 & 50 & 39 \\
KPNO 2.1m & T2KA     & 0.30 &  6 &  7 &  28 & 16 & 16 & 11 \\
FLWO 1.2m & AndyCam  & 0.32 & 40 &  - & 139 &  - & 40 &  6 \\
FLWO 1.2m & 4Shooter & 0.33 & 19 &  - & 145 & 29 & 34 & 33 \\
\hline
Total     &          &      & 65 &  7 & 391 & 74 &140 & 89 \\
\enddata
\label{tab:log}
\end{deluxetable}

\begin{deluxetable}{rrrcccccccl}
\tabletypesize{\footnotesize}
\tablewidth{0pc}
\tablecaption{Long period or non-periodic variables in NGC 6791}
\tablehead{\colhead{ID} & \colhead{$\alpha_{2000}[^h]$} &
\colhead{$\delta_{2000}[^\circ]$} &\colhead{$B_{max}$} &
\colhead{$V_{max}$} &\colhead{$I_{max}$} &\colhead{$A_B$} &
\colhead{$A_V$}& \colhead{$A_I$} & \colhead{$MP$ [\%]} & \colhead{Comments}}
\startdata
  B7 & 19.352056 & 37.799038 & 18.384 & 17.486 & 17.536 &  0.505 &  3.623 &  2.511 & 98      &  CV \\               
  B8 & 19.343262 & 37.747878 & 19.095 & 18.430 & 18.331 &  2.729 &  4.545 &  2.841 & \nodata &  CV \\               
 V10 & 19.353280 & 37.799508 & 20.469 & 19.578 & 18.328 &  0.786 &  0.402 &  0.273 & \nodata &  DEB$^*$\\           
 V13 & 19.347147 & 37.728627 & 15.617 & 13.986 & 10.704 &  0.051 &  0.171 &  0.330 & 98      &  Irr \\              
 V18 & 19.347051 & 37.769253 & 18.612 & 17.658 & 16.711 &  0.111 &  0.524 &  0.446 & 90      &  DEB \\              
 V19 & 19.347874 & 37.764148 & 16.715 & 15.099 & 13.936 &  0.155 &  0.249 &  0.200 & \nodata &  Irr \\              
 V20 & 19.348416 & 37.759655 & 18.295 & 17.295 & 16.332 &  0.251 &  0.377 &  0.234 & \nodata &  DEB \\              
 V21 & 19.349261 & 37.760280 & 18.404 & 17.465 & 16.567 &  0.052 &  0.139 &  0.072 & \nodata &  DEB$^*$ \\          
 V60 & 19.350193 & 37.762528 & 19.576 & 18.614 & 17.556 &  0.269 &  0.270 &  0.138 & 91      &  DEB$^*$ \\          
 V62 & 19.350848 & 37.731072 & 20.234 & 19.180 & 18.118 &  0.161 &  0.318 &  0.119 & \nodata &  \\                  
 V65 & 19.347909 & 37.791823 & 17.256 & 16.225 & 15.120 &  0.033 &  0.056 &  0.039 &  0      &  P$\sim11.3^d$?\\    
 V66 & 19.352340 & 37.748702 & 17.174 & 15.886 & 14.529 &  0.184 &  0.308 &  0.190 & \nodata &  P$\sim49^d,99^d$?\\ 
 V67 & 19.351021 & 37.801037 & 18.170 & 16.851 & 15.542 &  0.113 &  0.180 &  0.103 & 92      &  Irr\\               
 V70 & 19.342278 & 37.739207 & 16.330 & 14.777 & 10.581 &  0.295 &  0.397 &  0.277 & 62      &  Irr\\               
 V71 & 19.352913 & 37.723582 & 18.433 & 17.252 & 16.033 &  0.154 &  0.307 &  0.134 & 55      &  Irr\\               
 V72 & 19.351705 & 37.694531 & 15.925 & 15.111 & 14.230 &  0.029 &  0.045 &  0.025 & \nodata &  Irr\\               
 V73 & 19.348047 & 37.781267 & 15.728 & 14.833 & 13.871 &  0.036 &  0.080 &  0.051 & 20      &  P$\sim34.5^d$?\\    
 V74 & 19.352001 & 37.743069 & 15.921 & 14.654 & 13.353 &  0.025 &  0.037 &  0.060 & 91      &  Irr\\               
\enddata
\tablecomments{CV -- cataclysmic variable; DEB -- detached eclipsing binary;
Irr -- irregular; P -- period.\\
$^*$ No eclipses observed.}
\label{tab:misc}
\end{deluxetable}

\begin{deluxetable}{rrrlrrccccccc}
\tabletypesize{\footnotesize}
\tablewidth{0pc}
\tablecaption{Eclipsing binaries in NGC 6791}
\tablehead{\colhead{ID} & \colhead{$\alpha_{2000}[^h]$} &
\colhead{$\delta_{2000}[^\circ]$} &\colhead{P [d]} &\colhead{$\sigma_P$ [d]} 
&\colhead{$B_{max}$} &
\colhead{$V_{max}$} &\colhead{$I_{max}$} &\colhead{$A_B$} &
\colhead{$A_V$}& \colhead{$A_I$} & \colhead{$MP$ [\%]}}
\startdata
  V1 & 19.346558 & 37.742248 & ~0.2676758     & 1e-07 & 17.026 & 16.175 & 15.193 &  0.451 &  0.423 &  0.328 & \nodata \\
  V2 & 19.354874 & 37.766763 & ~0.2732310$^*$ & 8e-07 & 20.672 & 19.572 & 18.468 &  0.376 &  0.432 &  0.286 & \nodata \\
  V6 & 19.350752 & 37.813589 & ~0.2790259     & 2e-07 & 16.178 & 15.398 & 14.602 &  0.106 &  0.117 &  0.106 &  0      \\
  V5 & 19.346260 & 37.813300 & ~0.3126604     & 6e-07 & 18.062 & 17.167 & 16.250 &  0.107 &  0.073 &  0.113 & 98      \\
  V3 & 19.354377 & 37.769393 & ~0.3175443     & 9e-07 & 19.471 & 18.524 & 17.414 &  0.255 &  0.188 &  0.132 & 78      \\
  V4 & 19.348396 & 37.806629 & ~0.3255937     & 6e-07 & 18.689 & 17.727 & 16.725 &  0.136 &  0.139 &  0.128 & 98      \\
 V29 & 19.354796 & 37.751465 & ~0.436533$^*$  & 2e-06 & 20.971 & 19.757 & 18.340 &  0.483 &  1.024 &  0.329 & \nodata \\
  V7 & 19.340273 & 37.821882 & ~0.487558      & 7e-06 & 18.634 & 17.674 & 16.726 &  0.212 &  0.371 &  0.212 & \nodata \\
  B4 & 19.353583 & 37.764288 & ~0.796993      & 5e-06 & 17.805 & 17.812 & 17.999 &  0.075 &  0.089 &  0.142 &  4      \\
 V11 & 19.342581 & 37.804616 & ~0.883053      & 2e-06 & 20.386 & 19.349 & 18.295 &  0.502 &  0.803 &  0.428 & \nodata \\
 V12 & 19.345259 & 37.849029 & ~1.52321       & 2e-05 & 18.528 & 17.491 & 16.460 &  0.419 &  0.324 &  0.290 & 96      \\
 V31 & 19.350686 & 37.785924 & ~1.53617$^*$   & 3e-05 & 18.112 & 17.078 & 16.079 &  0.054 &  0.067 &  0.032 & 97      \\
 V32 & 19.341003 & 37.787297 & ~2.09576$^*$   & 7e-05 & 20.250 & 19.289 & 18.194 &  0.289 &  0.358 &  0.245 &  2      \\
 V33 & 19.344395 & 37.731804 & ~2.36626$^*$   & 4e-05 & 17.322 & 16.233 & 14.918 &  0.108 &  0.162 &  0.057 &  0      \\
  V9 & 19.346635 & 37.777072 & ~3.18670       & 4e-05 & 18.391 & 17.177 & 15.841 &  0.293 &  0.440 &  0.249 & 82      \\
 V37 & 19.355069 & 37.851982 & ~3.2199$^*$    & 1e-04 & 20.905 & 19.324 & 16.899 &  0.761 &  0.830 &  0.244 & \nodata \\
 V38 & 19.351025 & 37.768327 & ~3.90904       & 8e-05 & 19.732 & 18.713 & 17.663 &  0.293 &  0.327 &  0.168 & 92      \\
 V16 & 19.352108 & 37.802682 & ~4.4232        & 1e-04 & 18.707 & 17.745 & 16.772 &  0.186 &  0.173 &  0.095 & 96      \\
 V14 & 19.347685 & 37.756901 & 11.296$^*$     & 2e-03 & 19.540 & 18.557 & 17.526 &  0.209 &  0.172 &  0.105 &  0      \\
\enddata
\tablecomments{$^*$ Periods taken from M02.}
\label{tab:ecl}
\end{deluxetable}

\begin{deluxetable}{rrrlrccccccc}
\tabletypesize{\footnotesize}
\tablewidth{0pc}
\tablecaption{Other periodic variables in NGC 6791}
\tablehead{\colhead{ID} & \colhead{$\alpha_{2000}[^h]$} &
\colhead{$\delta_{2000}[^\circ]$} &\colhead{P [d]} & \colhead{$\sigma_P$ [d]}
&\colhead{$\langle B\rangle$} &
\colhead{$\langle V\rangle $} &\colhead{$ \langle I\rangle$} &\colhead{$A_B$} &
\colhead{$A_V$}& \colhead{$A_I$} & \colhead{$MP$ [\%]}}
\startdata
 V41 & 19.347492 & 37.806873 & ~0.481631     & 3e-06 & 20.285 & 19.244 & 17.708 &  0.059 &  0.087 &  0.052 & 77      \\
 V42 & 19.350057 & 37.714872 & ~0.506767$^*$ & 8e-06 & 20.679 & 19.631 & 18.336 &  0.018 &  0.061 &  0.042 & 92      \\
 V45 & 19.346127 & 37.701679 & ~4.578$^*$    & 1e-03 & 17.964 & 17.062 & 16.027 &  0.007 &  0.006 &  0.009 &  0      \\
 V46 & 19.355274 & 37.798951 & ~5.148$^*$    & 1e-03 & 19.777 & 18.624 & 17.273 &  0.073 &  0.027 &  0.020 & \nodata \\
 V48 & 19.352078 & 37.718526 & ~5.8278       & 6e-04 & 18.432 & 17.539 & 16.568 &  0.044 &  0.021 &  0.018 & 96      \\
 V17 & 19.344137 & 37.817930 & ~6.1752$^*$   & 9e-04 & 19.172 & 18.013 & 16.720 &  0.018 &  0.023 &  0.019 & 88      \\
 V51 & 19.353380 & 37.748572 & ~6.644$^*$    & 3e-03 & 21.072 & 20.003 & 18.708 &  0.051 &  0.017 &  0.009 & \nodata \\
 V52 & 19.355798 & 37.772039 & ~7.055$^*$    & 6e-03 & 18.329 & 17.497 & 16.591 &  0.012 &  0.008 &  0.004 & \nodata \\
 V53 & 19.350232 & 37.743193 & ~7.470$^*$    & 3e-03 & 19.657 & 18.781 & 17.822 &  0.019 &  0.009 &  0.006 & 86      \\
 V54 & 19.355196 & 37.726818 & ~8.333        & 1e-03 & 17.461 & 16.485 & 15.441 &  0.022 &  0.017 &  0.009 & \nodata \\
 V56 & 19.345907 & 37.763587 & 11.863        & 3e-03 & 17.989 & 17.042 & 16.070 &  0.011 &  0.009 &  0.005 & 98      \\
 V58 & 19.354038 & 37.801260 & 13.007        & 2e-03 & 18.362 & 17.520 & 16.599 &  0.009 &  0.016 &  0.009 & 87      \\
 V68 & 19.341891 & 37.848689 & 15.26         & 1e-02 & 14.786 & 14.211 & 13.834 &  0.033 &  0.003 &  0.031 & \nodata \\
 V69 & 19.354467 & 37.779617 & 43.26         & 5e-02 & 15.357 & 13.844 & 10.650 &  0.011 &  0.055 &  0.023 & 47      \\
\enddata
\tablecomments{$^*$ Periods taken from M02.}
\label{tab:pul}
\end{deluxetable}

\end{document}